\documentclass[nofootinbib]{revtex4-2}

\usepackage{geometry}                
\usepackage{amsmath,amssymb,amsfonts,dcolumn,color,graphicx}
\usepackage{latexsym,placeins,epsfig,tikz,amsbsy,bm,enumitem}
\usetikzlibrary{arrows,shapes}
\usepackage{subfigure,rotating,bm,mathrsfs}
\usepackage[title]{appendix}
\usepackage[pagebackref=false, colorlinks=true]{hyperref}
\hypersetup{linkcolor=red,         
                  citecolor=blue,        
                  filecolor=magenta,      
                  urlcolor=blue}          

\newcommand{\calH}{{\cal H}}

\allowdisplaybreaks

\begin{document}

\title{\Large Cosmological scalar and tensor perturbations with a scalar field: quadratic-order effective energy-momentum tensor}

\author{Inyong Cho}
\email{iycho@seoultech.ac.kr}
\affiliation{School of Natural Sciences, College of Liberal Arts,
Seoul National University of Science and Technology, Seoul 01811, Korea} 

\begin{abstract}
We introduce the scalar and tensor modes of the gravitational perturbation
in the presence of a scalar field which describes inflation.
We investigate the back-reaction of the perturbations to the background
by studying the effective energy-momentum tensor (2EMT) which is the second order
constructed by the quadratic terms of the linear perturbations.
2EMT is gauge dependent due to the scalar mode.
We obtain 2EMT in the slow-roll stage of inflation,
and get its cosmological expressions in three 
(longitudinal, spatially flat, and comoving) gauge conditions.
We find that the pure scalar-mode part in 2EMT is stronger in the short-wavelength limit, 
while the parts involved with the tensor mode 
(the pure tensor-mode part and the scalar-tensor coupled part)
are stronger in the long-wavelength limit.
\end{abstract}

\maketitle


\section{Introduction}

For the last decades, the cosmological perturbation theory has been well established,
and applied to cosmological situations. 
In particular, the density perturbation during inflationary period could explain
the data of the cosmic microwave background radiation successfully.
As the observational precision is improved, 
people starts to pay more attention to the higher-order phenomena than linear.

Recently the effective energy-momentum tensor (2EMT) produced by a fluid \cite{Cho:2020zbh} 
and by a scalar field \cite{Cho:2022maa} has been investigated individually.
2EMT was constructed by the quadratic terms of the linear perturbations of the matter and the gravitational fields.
The gauge issue of 2EMT was studied in those works.
2EMT was found to be gauge dependent in both works.
The gauge dependence of 2EMT of a scalar field had been issued in Refs.~\cite{Mukhanov:1996ak,Abramo:1997hu} 
and in Refs.~\cite{Unruh:1998ic,Ishibashi:2005sj,Green:2010qy}.
More work have been done in Refs.~\cite{Geshnizjani:2002wp,Brandenberger:2002sk,Geshnizjani:2003cn,Martineau:2005aa}.
In all of these works, only the scalar perturbation of the gravitational field 
had been considered. 

In this work, we investigate 2EMT of a scalar field in the inflationary period.
We consider both of the scalar and the tensor perturbations together.
In the linear order, they are decoupled and the solutions are the same with
those obtained separately.
In the second order, however, the quadratic-order terms contain the coupled terms
of the scalar and the tensor modes.

In scalar-vector-tensor (SVT) decomposition of the gravitational perturbations,
the scalar or tensor mode of perturbations can be considered solely 
with the other modes turned off in {\it vacuum}.
However, when a ``matter field" is introduced,
the tensor mode cannot be considered solely while the scalar mode can.
It is because the tensor mode represents only {\it two} degrees of freedom 
which are responsible for the gravity wave
while the matter-field degree of freedom is implied by the scalar mode.
Therefore, if there is a matter field, 
we need to consider the tensor mode together with the scalar mode,
which inevitably induces the coupled scalar-tensor terms in the quadratic order.

In this work, we shall consider the cosmological perturbations during inflation.
We introduce the matter-field perturbation of the inflaton (scalar field)
and the scalar and the tensor modes of the gravitational perturbations.
We shall obtained 2EMT constructed by the quadratic terms of the linear-order perturbations. 
It contains the scalar-only terms that were obtained in Refs.~\cite{Cho:2022maa,Mukhanov:1996ak,Abramo:1997hu},
and the tensor-only terms obtained in Refs.~\cite{Mukhanov:1996ak,Abramo:1997hu}.
In addition there are the coupled terms of scalar and tensor modes.

2EMT is constructed as following.
We expand the Einstein's equation by order,
then the second-order equation can be written as
\begin{equation}\label{GmuTmu}
G_{\mu\nu}^{(1)}[g^{(2)}] + G_{\mu\nu}^{(2)}[g^{(1)}]
= 8\pi G T_{\mu\nu}^{(1)}[g^{(2)},\delta\phi^{(2)}]
+ 8\pi G T_{\mu\nu}^{(2)}[g^{(1)},\delta\phi^{(1)}].
\end{equation}
This equation consists of two parts;
the linear-order terms of the second-order perturbations (the first term of each side),
and the quadratic-order terms of the first-order perturbations (the second term of each side).
Here, $g$ and $\delta\phi$ stand for the metric and the matter perturbations individually.
2EMT is made of the quadratic terms,
\begin{equation}\label{T2eff}
T^{(2,{\rm  eff})}_{\mu\nu} \equiv 
T_{\mu\nu}^{(2)}[g^{(1)},\delta\phi^{(1)}] - \frac{G_{\mu\nu}^{(2)}[g^{(1)}]}{8\pi G} .
\end{equation}
This is regarded as the source of the back-reaction to the geometry part 
$G_{\mu\nu}^{(1)}[g^{(2)}]$ in Eq.~\eqref{GmuTmu},
while the linear-order term $T_{\mu\nu}^{(1)}[g^{(2)},\delta\phi^{(2)}]$ has 
stochastically vanishing effect when it is integrated in time \cite{Mukhanov:1996ak,Abramo:1997hu}.

The paper is organized as follows.
In Sec.~\ref{SecPert}, we introduce the linear perturbations and their solutions,
in particular in the slow-roll regime of inflation.
In Sec.~\ref{Sec2EMT}, we evaluate 2EMT in the long- and short-wavelength limits
with three (longitudinal, spatially flat, and comoving) gauge conditions 
imposed on the scalar perturbation.
We present 2EMT of the {\it scalar-only}, the {\it tensor-only}, and the {\it scalar-tensor coupled} terms individually,
and discuss its significances.
We conclude in Sec.~\ref{Conc}.

\section{Perturbations and linear solutions}
\label{SecPert}

\subsection{Perturbations}

Let us consider the general metric 
\begin{align}\label{metric1}
ds^{2} 
= a^{2}(\eta) \Big[ -(1+2A)d\eta^{2} - 2 B_{i} d\eta dx^{i}
+ (\delta_{ij} + 2 C_{ij}) dx^{i} dx^{j} \Big],
\end{align}
where the metric perturbations $A$, $B_i$, and $C_{ij}$ represent all orders.
We consider the scalar and the tensor modes of the metric perturbations,
\begin{align}
A = \alpha   ,
\quad
B_i = \beta_{,i}   ,
\quad
C_{ij} = -\psi\delta_{ij} + E_{,ij} +h_{ij},
\end{align}
where four functions, $\alpha$, $\beta$, $\psi$, and $E$ are the scalar modes,
and $h_{ij}$ is the tensor mode.

The energy-momentum tensor of scalar field (inflaton) $\phi$ is given by 
\begin{align}\label{emt1}
T_{\mu\nu} 
=\phi_{,\mu}\phi_{,\nu} - g_{\mu\nu} \bigg[ \frac{1}{2} g^{\rho\sigma} \phi_{,\rho} \phi_{,\sigma} + V(\phi) \bigg] ,
\end{align}
where $V(\phi)$ is the potential. 
The scalar field $\phi$ is the sum of the background field $\phi_0$ 
and the perturbation $\delta\phi$,
\begin{align}
\phi = \phi_0 + \delta\phi   .
\end{align}        

With Eqs.~\eqref{metric1} and \eqref{emt1}, up to the second order in perturbations,
the Einstein tensor $G_{\mu\nu}$ was presented in Eqs.~(2)-(5) of Ref.~\cite{Cho:2020zbh},
and the energy-momentum tensor was presented in Eqs.~(10)-(12) of Ref.~\cite{Cho:2022maa}.

The scalar-mode and the tensor-mode equations are decoupled in the linear-order Einstein's equation.
For the scalar perturbations, let us define the gauge invariant variables up to the first order,
\begin{align}
\overline{\delta \phi} & = \delta \phi - \phi_{0}^{\prime}  Q   ,\label{GIdeltaphi}\\
\Phi & = \alpha - Q^{\prime} - \mathcal{H} Q   ,\\
\Psi & = \psi + \mathcal{H} Q   ,
\end{align} 
where $\mathcal{H} =a'/a$ with $'\equiv d/d\eta$, and 
\begin{equation}
Q = \beta + E'.
\end{equation}
Then, the first-order equations are singled out using the Einstein's equation and the scalar-field equation  
as a gauge-invariant form,
\begin{align}
\label{eq:Psi-eq}
\Psi^{\prime \prime} - \Delta \Psi + 2 K \Psi^{\prime} + 2L \Psi = 0,
\end{align}
where
\begin{align}
K = 3\mathcal{H} + a^{2} \frac{V_{\phi} }{ \phi_{0}^{\prime} } ,\quad
L = \mathcal{H}^{\prime} + 2\mathcal{H}^{2} 
+a^{2} \mathcal{H} \frac{V_{\phi} }{ \phi_{0}^{\prime}},
\end{align}
where $V_{\phi}=dV/d\phi$.
Here, we used the relations obtained from the first-order equations for the gauge-invariant variables, 
$\Psi = \Phi$, and
\begin{align}\label{GIdeltaphi2}
\overline{\delta \phi} = \frac{\Psi^{\prime} + \mathcal{H} \Psi}{4\pi G \phi_{0}^{\prime}},\quad
\overline{\delta \phi}^{\prime} = \frac{\Delta \Psi - K \Psi^{\prime} - L \Psi }{4\pi G \phi_{0}^{\prime}}.
\end{align}

For the tensor perturbation $h_{ij}$, we impose transverse-traceless (TT) gauge  
to keep the pure tensor terms in $C_{ij}$ in Eq.~\eqref{metric1} 
(excluding the contributions from the derivative terms of the scalar and vector modes)
\cite{Mukhanov:1990me}.
The first-order equation for $h_{ij}$ is given by 
\begin{align}\label{eq:h}
h''_{ij}+2 \mathcal{H} h'_{ij}- \Delta h_{ij} = 0.
\end{align}

We introduce the Fourier-modes for $\Psi$ and $h_{ij}$ 
which will be more convenient in studying perturbations,
\begin{align}
\Psi (\eta,{\bf x}) &=  \int d^3{\bf k} {\Psi}_{\bf k}(\eta) e^{i {\bf k} \cdot {\bf x}},\label{eq:FR1}\\
h_{ij} (\eta,{\bf x}) &= \epsilon^\lambda_{ij} \int d^3{\bf k} h_{\bf k}(\eta) e^{i {\bf k} \cdot {\bf x}},\label{eq:FR2}
\end{align}
where $\epsilon^\lambda_{ij}$ is the polarization tensor.

\subsection{Linear solutions}
Let us consider the inflationary stage in the slow-roll regime. 
We define the slow-roll parameters as
\begin{align}
\epsilon & \equiv -\frac{\dot{H}}{H^2} = 4\pi G \frac{\dot\phi_0^2}{H^2} ,\\
\delta & \equiv -\frac{\ddot\phi_0}{H\dot\phi_0} = \epsilon - \frac{\dot\epsilon}{2H\epsilon} ,
\end{align}
where the overdot represents the derivative with respect to the cosmological time $t$,  
$\dot{} \equiv d/dt = (1/a)d/d\eta$,
and $H=\dot{a}/a$.

For the scalar perturbation, the solution becomes \cite{Mukhanov:1990me}
\begin{align}\label{eq:Psi-sol}
\Psi_{\bf k}(\eta) \approx \left\{
\begin{array}{ll}
A_1\epsilon & \quad (\text{for long wavelength})
\vspace{0.5em}
\\
4\pi G \dot{\phi}_0 [c_1\sin(k\eta) + c_2 \cos(k\eta)]
& \quad (\text{for short wavelength})
\end{array}
\right.,
\end{align}
where $A_1$ and $c_i$ are complex constants,
and $k = \sqrt{{\bf k}\cdot{\bf k}}$.
Note that $\Psi_{\bf k}$ is approximately constant in the long-wavelength limit 
and rapidly oscillates in the short-wavelength limit.
The long-wavelength limit corresponds to $k \ll \cal H$ for which
the physical wavelength $\lambda_\text{phys} =a/k$ of the perturbation modes 
is much larger than the Hubble scale $H^{-1}$, $\lambda_{\rm phys} \gg H^{-1}$.
In the inflationary stage, the scale factor is approximated by $a(\eta) \approx -1/(H\eta)$,
where $\eta<0$.
Then one gets ${\cal H} \approx -1/\eta$, 
and the condition for the long-wavelength limit  becomes $k \ll {\cal H} \to -k\eta \ll 1$.
The short-wavelength limit corresponds to $k \gg \calH$ 
which reduces to $-k\eta \gg 1$ in the inflationary stage.

For the tensor perturbation, we define a new variable as 
\begin{align}
v_{\bf k}(\eta)=\frac{a(\eta)h_{\bf k}(\eta)}{\sqrt{16\pi G}},
\end{align}
then Eq.~\eqref{eq:h} becomes
\begin{align}
v_{\bf k}'' +  \left(k^2-\frac{\mu ^2-1/4}{\eta ^2}\right)v_{\bf k}=0,
\quad \mu^2=\frac{9}{4}+3\epsilon+\cdots.
\end{align}
The solution to this equation is given by the spherical Bessel functions as
\begin{align}\label{solv}
v_{\bf k}(\eta) = \sqrt{-\eta } \left[ b_1 J_\mu(-k \eta) +b_2Y_\mu(-k \eta ) \right],
\end{align}
where $b_1$ and $b_2$ are complex constants.\footnote{In this work, 
we consider the solutions in a general form with keeping the constants in Eqs.~\eqref{eq:Psi-sol} and \eqref{solv}.
However, if we impose the initial conditions of the Bunch-Davies vacuum for inflation,
the constants have the relations, $b_1=-ib_2$ and $c_1=-ic_2$.
The normalization condition would further determine these constants as a function of $k$.}

We perform expansions of this solution in the slow-roll parameter $\epsilon$ in the index $\mu$,
At the same time, we need to perform additional expansions in the wave lengths in the argument,
\begin{align}
\sigma_L &\equiv -k\eta \ll 1 \quad \mbox{(for long wavelength)},\label{sigmaL}\\
\sigma_S &\equiv -(k\eta)^{-1} \ll 1 \quad \mbox{(for short wavelength)}.\label{sigmaS}
\end{align}
Then in the long-wavelength limit, we get  
\begin{align}
h_{\bf k}(\eta) = &\frac{\sqrt{2G}}{9k^{3/2}\eta a} 
\left\{
12b_1k^3\eta ^3 +18b_2k^2\eta^2 
\left[ 1 -\epsilon\log(-k\eta) +\epsilon \left( -2 +\log2 +\psi^{(0)}(3/2)\right)\right] \right. \nonumber\\
&+b_2 \left[ 36 -6\epsilon^3\log(-k\eta)^3 +36\epsilon \left( \log2 +\psi^{(0)}(3/2)\right) \right. \nonumber\\
&+3\epsilon^2 \left( -24 +3\pi^2 +6\log2^2 -4\psi^{(0)}(3/2)
+6 \psi^{(0)}(3/2)^2 +4\log2 \left( -1 +3\psi^{(0)}(3/2) \right) \right) \nonumber\\
&+3\epsilon^2 \log(-k\eta)^2 \left[ 6+ \epsilon  \left(-4 +6\log2 +6 \psi^{(0)}(3/2) \right)\right] \nonumber\\
&-\epsilon\log(-k\eta) \left[ 36 +12\epsilon \left( -1 +3\log2 +3\psi^{(0)}(3/2)\right) \right. \nonumber\\
&\left. +\epsilon^2 \left( 9\pi^2 -64-24\log2 +18\log^22 -24\psi^{(0)}(3/2) +36\log2 \psi^{(0)}(3/2) +18\psi^{(0)}(3/2)^2\right) \right]  \nonumber\\
&+\epsilon^3 \left[
48 -64\log2 -6\log^22
+\left( -64 -24\log2 +18\log^22 \right) \psi^{(0)}(3/2) \right.  \nonumber\\
&\left.\left. +(-12 +18\log2) \psi^{(0)}(3/2)^2
+6\psi^{(0)}(3/2)^3
+6\psi^{(2)}(3/2 )\right]  \right\} 
+\cdots,
\label{hL}
\end{align}
where $\psi^{(n)}(z) = d^n\psi(z)/dz^n = d^{n+1}\log\Gamma(z)/dz^{n+1}$.
The solution has been expanded in the small quantities such as the slow-roll parameter $\epsilon$ 
and the wave-length parameter $k\eta = -\sigma_L$ in the order of ${\cal O}(\varepsilon^{-1\sim 2})$
where $\varepsilon = \epsilon$, or $\sigma_L$. 
Afterwards the expression has been recovered, $\sigma_L \to -k\eta$.

In the short-wavelength limit, we get  
\begin{align}
h_{\bf k}(\eta) = &\frac{\sqrt{G}}{432\sqrt{2}k^{7/2}\eta^3 a}
\left\{
1728\epsilon {\cal SC} 
-1296\epsilon k\eta  \left[
( -(2+3\epsilon){\cal CS} + \pi\epsilon {\cal SC}  \right] \right. \nonumber\\
&-24k^2\eta^2 \left[
\pi\epsilon  (-72 -84\epsilon +(20+3\pi^2)\epsilon^2) {\cal CS} 
\left. +3(-48 -72\epsilon +6\pi^2\epsilon^2 +5\pi^2\epsilon^3){\cal SC} \right] \right. \nonumber\\
&-k^3\eta^3 \left. \left[
9(384 -48\pi^2\epsilon^2 +32\pi^2\epsilon^3 +\pi^2(\pi^2-80/3)\epsilon^4){\cal CS} 
-8\pi\epsilon( +216 -72\epsilon -3(3\pi^2-16)\epsilon^2 +(9\pi^2-40)\epsilon^3){\cal SC}  \right] \right\}
+\cdots,
\label{hS}
\end{align}
where ${\cal SC} \equiv b_1\sin(k\eta)+b_2\cos(k\eta)$ and ${\cal CS} \equiv b_1\cos(k\eta) -b_2\sin(k\eta)$.
The solution has been expanded in the small quantities $\epsilon$ 
and the wave-length parameter $(k\eta)^{-1} =-\sigma_S$ in the order of ${\cal O}(\varepsilon^{0\sim 4})$
where $\varepsilon = \epsilon$, or $\sigma_S$. 
Afterwards the expression has been recovered, $\sigma_S \to -(k\eta)^{-1}$.

\section{Second-order effective energy-momentum tensor (2EMT)}
\label{Sec2EMT}

When both of the scalar and the tensor modes are turned on, 
2EMT consists of three parts;
the scalar-mode only ($\cal{S}$), 
the tensor-mode only ($\cal{T}$), 
and the scalar-tensor coupled ($\cal{ST}$) parts. 
In this section, we define a new notation for 2EMT for convenience as
\begin{align}
\hat\tau_{\mu\nu} \equiv  8\pi G \tau_{\mu\nu}({\bf k}) 
= 8\pi G \left\langle T_{\mu\nu}^{(2,\text{eff})}({\bf k}) \right\rangle,
\end{align}
where $\tau_{\mu\nu}({\bf k})$ means $\tau_{\mu\nu}$ for a given ${\bf k}$ mode,
and $\left\langle \right\rangle$ denotes the ``Fourier-transformed" quantity
with the Fourier modes introduced in Eqs.~\eqref{eq:FR1} and ~\eqref{eq:FR2}.
We expand $\hat\tau_{\mu\nu}$ in terms of the {\it small parameter}  $\varepsilon$ 
such as the slow-roll parameters $\epsilon$ and $\delta$, 
and the wave-length parameters $\sigma_L$ and $\sigma_S$.

\vspace{12pt}
\noindent
\underline{Summary of result}
\vspace{12pt}

\noindent 
1. Long-wavelength limit: $\cal{T}$ and $\cal{ST}$ parts are dominant.\footnote{The comoving gauge 
exhibits somewhat peculiar features in the overall results.
For example, $\hat\tau_{\mu\nu}$ for $\cal{S}$ part of 2EMT in the comoving gauge
is equally dominant to those for $\cal{T}$ and $\cal{ST}$ parts, 
while remaining subdominant in the other gauges.
We shall not focus much on the result of comoving gauge in this work.
We shall come back to this peculiar feature of comoving gauge in the future
after investigating more gauge conditions.}

\vspace{12pt}
(i) $00$-component: $\cal{T}$ part is dominant. 

(ii) $ij$-component: $\cal{T}$ part is the most dominant.
$\cal{ST}$ part is the next order along with $\cal{T}$.

\vspace{12pt}
\noindent 
2. Short-wavelength limit: $\cal{S}$ part is mainly dominant,
while $\cal{T}$ part is of the same order
for the $00$- and $33$-components.\footnote{This statement is generally true {\it excluding the comoving gauge}.}

\subsection{Scalar-only ($\cal{S}$)}\label{SecScalar}

2EMT for $\cal{S}$ terms 
was investigated for a fluid in Ref.~\cite{Cho:2020zbh},
and for a scalar field in Ref.~\cite{Cho:2022maa}.
A comprehensive view of 2EMT across different (long- and short-) wavelength limits 
and three (longitudinal, spatially-flat, comoving) gauge choices have been investigated.
We do not duplicate the results here,
and readers refer those references.
Instead, we present here the summary of the order dependence of $\hat\tau_{\mu\nu}$ on $\varepsilon$
for the slow-roll regime of inflation investigated in Ref.~\cite{Cho:2022maa}.

\subsubsection{Long-wavelength limit}

2EMT in the long-wavelength limit for longitudinal (LG), spatially-flat (SF), and comoving (CM) gauge
has been obtained in Eqs.~(46), (47), (60), (61), (73), and (74) in Ref.~\cite{Cho:2022maa}.
The results can be expressed in terms of $\varepsilon$ as
\begin{align}
\hat{\tau}^\text{LG}_{\rm 00} &\propto \hat{\tau}^\text{LG}_{\rm ii} \sim {\cal O}(\varepsilon^{0}) , \\
\hat{\tau}^\text{SF}_{\rm 00} &\propto \hat{\tau}^\text{SF}_{\rm ii} \sim {\cal O}(\varepsilon^{0}) , \\
\hat{\tau}^\text{CM}_{\rm 00} &\propto \hat{\tau}^\text{CM}_{\rm ii} \sim {\cal O}(\varepsilon^{-2}+\varepsilon^{-1}) .
\end{align}
Here, ${\cal O}(\varepsilon^{-2})$ comes from $\sigma_L$ contribution only in Eq.~\eqref{sigmaL},
while the others contain the slow-roll contributions as well.

\subsubsection{Short-wavelength limit}

2EMT in the short-wavelength limit
has been obtained in Eqs.~(52), (53), (64), (65), (77), and (78) in Ref.~\cite{Cho:2022maa}.
The results can be expressed in terms of $\varepsilon$ as
\begin{align}
\hat{\tau}^\text{LG}_{\rm 00} &\propto \hat{\tau}^\text{LG}_{\rm ii} \sim {\cal O}(\varepsilon^{0}+\varepsilon^{3}) , \\
\hat{\tau}^\text{SF}_{\rm 00} &\propto \hat{\tau}^\text{SF}_{\rm ii} \sim {\cal O}(\varepsilon^{0}+\varepsilon^{3}) , \\
\hat{\tau}^\text{CM}_{\rm 00} &\propto \hat{\tau}^\text{CM}_{\rm ii} \sim {\cal O}(\varepsilon^{-1}) .
\end{align}
Here, ${\cal O}(\varepsilon^{0})$ contains no small parameters ($\varepsilon$),
while the others come from the slow-roll contributions.
This ${\cal O}(\varepsilon^{0})$ is the strongest order in the short-wavelength limit of 2EMT.

\subsection{Tensor-only ($\cal{T}$)}\label{SecTensor}

2EMT for $\cal{T}$ terms as presented in  Refs.~\cite{Mukhanov:1996ak,Abramo:1997hu} 
is rewritten as
\begin{align}
\hat{\tau}_{00} &= -\frac{3}{2} \left(h_{ij,{n}}\right){}^2-2 h_{ij} h_{ij,{n}{n}}+4 \mathcal{H} h_{ij} h'_{ij}+\frac{1}{2}
(h'_{ij})^2,\\
\hat{\tau}_{ij} &= -2 h_{i{n},{m}} h_{j{n},{m}}+h_{{n}{m},i} h_{{n}{m},j}+2 h'_{i{n}} h'_{j{n}}
+\delta _{ij} \left[\frac{3}{2} \left(h_{{n}{m},{n3}}\right){}^2-\frac{3}{2}
(h'_{{n}{m}})^2\right].
\end{align}
For the tensor mode, let us consider the wave propagating \textbf{\textit{along $x^3$ direction}},
i.e., $k=k_3$.
Using the solutions \eqref{hL} and \eqref{hS}, 
we evaluate $\hat{\tau}_{\mu\nu}$ in the long- and short-wavelength limits.

\subsubsection{Long-wavelength limit}

The components of 2EMT in the small parameter $\varepsilon$
are given by orders as following [in this limit, $k\eta$ $(=-\sigma_L)$ in the expression is regarded small];

\begin{align}
\hat{\tau}_{00} &= 
-\frac{224 G|b_2|^2}{k\eta^2a^2 }-\frac{256 G {\rm Re}(b_1b_2^*)}{\eta a^2}
-\frac{256 G|b_2|^2 \epsilon^2}{3 k^3 \eta ^4 a^2}
+ \frac{64G|b_2|^2 \epsilon }{k\eta ^2a^2}
\left[ 7\log(-k\eta) +4 -7\log2 -7\psi^{(0)}(3/2) \right] \nonumber\\
&+\frac{64G|b_2|^2\epsilon^3}{9k^3\eta^4 a^2} 
\left[
24 \log(-k\eta)
-46 +51\log2 -33\log^22
+\left( -102 +87\log2 -9\log^22 \right)\psi^{(0)}(3/2)
-9 \left( 1 -\log2 \right)\psi^{(0)}(3/2)^2
\right]
\cdots \\
& \sim {\cal O}(\varepsilon^{-2}+\varepsilon^{-1}).\nonumber
\end{align}
Here, the first two terms are the $\sigma_L$ contributions in the order of 
${\cal O}(\varepsilon^{-2})$ and ${\cal O}(\varepsilon^{-1})$.
[For the Bunch-Davies initial conditions, the second term vanishes, ${\rm Re}(b_1b_2^*)=0$.] 
The remaining terms contain the slow-roll contributions as well, 
in the order of 
${\cal O}(\varepsilon^{-2})$ for the third term and ${\cal O}(\varepsilon^{-1})$ for the rests.\footnote{Note that
the logarithmic divergence near $k\eta =0$, i.e., $\log\varepsilon$ is weaker 
than that of $1/\varepsilon$.}
The spatial components are given by
\begin{align}
\hat{\tau}_{11} = \hat{\tau}_{22}  
= \frac{32G|b_2|^2}{k\eta^2a^2} 
 \left\{ 1 + 2\epsilon  \left[ -\log(-k\eta) +\log2 + \psi^{(0)}(3/2) \right] \right\}
 +\cdots \sim {\cal O}(\varepsilon^{-2}+\varepsilon^{-1}),
\end{align}
\begin{align}
\hat{\tau}_{33} 
=\frac{160G|b_2|^2}{k\eta^2a^2}
 \left\{ 1 + 2\epsilon  \left[-\log(-k\eta) +\log2 +\psi^{(0)}(3/2)\right] \right\}
 +\cdots
\; \approx 5\hat{\tau}_{11}.
\end{align}
Here, the first term is the $\sigma_L$ contribution of ${\cal O}(\varepsilon^{-2})$,
and the second term of ${\cal O}(\varepsilon^{-1})$ contains the slow-roll contributions as well.
${\cal O}(\varepsilon^{-2})$ is the strongest order in the long-wavelength limit of 2EMT.

\subsubsection{Short-wavelength limit}

For the short-wavelength limit, 
we take the temporal average over a period for the sinusoidal functions coming from Eq.~\eqref{hS}.
Then 2EMT becomes [in this limit, $(k\eta)^{-1}$ $(=-\sigma_S)$ in the expression is regarded small]
\begin{align}
\hat{\tau}_{00} &= \frac{16G k (|b_1|^2+|b_2|^2)}{a^2} 
\left[ 2- \frac{7(1+2\epsilon)}{k^2\eta^2} \right] +\cdots \sim {\cal O}(\varepsilon^{0}+\varepsilon^{2}+\varepsilon^{3}),\\
\hat{\tau}_{11} &= \hat{\tau}_{22} =  \frac{16G (|b_1|^2+|b_2|^2)}{k\eta^2a^2}(1+\epsilon) +\cdots \sim {\cal O}(\varepsilon^{2}+\varepsilon^{3}) ,\\
\hat{\tau}_{33} &= \frac{16G k (|b_1|^2+|b_2|^2)}{a^2} 
\left[ 2+ \frac{5+6\epsilon}{k^2\eta^2} \right] +\cdots \sim {\cal O}(\varepsilon^{0}+\varepsilon^{2}+\varepsilon^{3}).
\end{align}
The terms of ${\cal O}(\varepsilon^{0})$ contain no small parameters ($\varepsilon$).
The terms of ${\cal O}(\varepsilon^{2})$ are from the $\sigma_S$ contribution,
and the terms of ${\cal O}(\varepsilon^{3})$ contain the slow-roll contribution as well. 
$\hat{\tau}_{00}$ and $\hat{\tau}_{33}$ of $\cal{T}$
have the same (strongest) order ${\cal O}(\varepsilon^{0})$
with that of $\cal{S}$ in this lint.
(For the Bunch-Davies initial conditions, the coefficients become $|b_1|^2+|b_2|^2 = 2|b_2|^2$.)

\subsection{Scalar-Tensor coupled ($\cal{ST}$)}

2EMT for $\cal{ST}$ terms is given by
\begin{align}
\hat{\tau}_{ij} &=h'_{ij} \left( Q'' +2\mathcal{H}Q' +2\mathcal{H}'Q +\Delta Q-2 \Delta E' \right)\nonumber\\
&+ h_{ij} \left\{ 4\left( \mathcal{H} +a^2\frac{V_\phi}{\phi_0'} \right)\Psi ' 
+4\left( \mathcal{H}' +2\mathcal{H}^2 +a^2\mathcal{H}\frac{V_\phi}{\phi_0'}\right)\Psi \right.\nonumber\\
&+2\mathcal{H}Q'' +4\left( \mathcal{H}'+\mathcal{H}^2\right)Q' 
+2\Delta Q' +2 \mathcal{H}\Delta Q  
-2\left[3\mathcal{H}\mathcal{H}' +2\mathcal{H}^3 -4\pi Ga^2 (3\mathcal{H}V_0 +2\phi_0'V_\phi)\right]Q \nonumber\\
&\left. -\Delta E'' -2 \mathcal{H} \Delta E' -\Delta^2 E \right\}  .
\end{align}
For 2EMT of $\cal{ST}$ terms,
the nonzero components are 
$\hat{\tau}_{11} = -\hat{\tau}_{22} \equiv \hat{\tau}_{\rm ST}$ for $+$-polarization,
and $\hat{\tau}_{12} = \hat{\tau}_{21} = \hat{\tau}_{\rm ST}$ for $\times$-polarization of the tensor mode.
The contributions to $\hat{\tau}_{ij}$ are from the geometric part $G_{\mu\nu}^{(2)}[g^{(1)}]$ and
the matter part $T_{\mu\nu}^{(2)}[g^{(1)},\delta\phi^{(1)}]$ in Eq.~\eqref{T2eff}.
The matter-part contribution is given by
\begin{align}
\hat{\tau}_{ij}\left[T_{\mu\nu}^{(2)}\right] = 
h_{ij} \left\{ 4\Delta\Psi - 4\left( 3\mathcal{H} +2a^2\frac{V_\phi}{\phi_0'} \right)\Psi '
-4\left( 3\mathcal{H}^2 +2a^2\mathcal{H}\frac{V_\phi}{\phi_0'}\right)\Psi
+4\left(3\mathcal{H}\mathcal{H}' -3\mathcal{H}^3 -8\pi Ga^2\phi_0'V_\phi)\right)Q \right\}.
\end{align}

We impose three gauge conditions; 
\\

(i) longitudinal gauge: $\beta=E=0$

(ii) spatially-flat gauge: $\psi=E=0$ $\quad\Rightarrow\quad$  $Q=\Psi/{\cal H}$

(iii) comoving gauge: $\delta\phi=E=0$ $\quad\Rightarrow\quad$ $\Psi=\psi +{\cal H}\beta$ and $Q=(\Psi'+{\cal H}\Psi)/({\cal H}'-{\cal H}^2)$
\\

\noindent
For (i) and (ii), the results depend on the slow-roll parameters $\epsilon$ and $\delta$.
2EMT vanishes in the pure-de Sitter limit.
For (iii), however, the results exhibit peculiar features as in $\cal{S}$ case in Sec.~\ref{SecScalar},
for example, the inverse proportionality to the slow-roll parameters.

\subsubsection{Long-wavelength limit}

2EMT in the long-wavelength limit with three gauge conditions are obtained as
\begin{align}
\hat{\tau}^\text{LG}_{\cal ST} &= -\frac{8 \sqrt{2} \sqrt{G} \mathcal{H}^2 (A_1b_2^*+A_1^*b_2)}{k^{3/2}\eta a}  \epsilon(3\epsilon -2\delta) +\cdots 
\quad\sim {\cal O} (\varepsilon^{-1}) ,\\
\hat{\tau}^\text{SF}_{\cal ST} &= -\frac{2 \sqrt{2} \sqrt{G} \mathcal{H}\left(1-3 \mathcal{H} \eta\right) \left(A_1b_2^* +A_1^*b_2\right)}{k^{3/2}\eta^2 a} \epsilon ^2 +\cdots 
\quad\sim {\cal O} (\varepsilon^{-1}) ,\\
\hat{\tau}^\text{CM}_{\cal ST} &= -\frac{8 \sqrt{2} \sqrt{G} \mathcal{H} \left(A_1b_2^*+A_1^* b_2\right)}{k^{3/2}\eta^2 a} +\cdots 
\quad\sim {\cal O} (\varepsilon^{-3}) .
\end{align}
Here, $k/{\cal H} = -k\eta$ $(=\sigma_L)$ in the expression is regarded small.
The slow-roll correction is of the same order 
with that in $\hat{\tau}_{ii}$ of $\cal{T}$ in Sec.~\ref{SecTensor}.
[However, the $\sigma_L$ contribution of $\cal{T}$ is stronger, $\sim {\cal O} (\varepsilon^{-2})$.]

\subsubsection{Short-wavelength limit}

After taking the temporal average for the sinusoidal functions, 2EMT becomes
\begin{align}
\hat{\tau}^\text{LG}_{\cal ST} &= -\frac{8\sqrt{2 \pi } G  \mathcal{H}^2 k^{1/2} \left(c_1^* b_1+c_2^* b_2+b_1^* c_1+b_2^* c_2\right)}{a^2} \sqrt{\epsilon }  +\cdots  
\quad\sim {\cal O} (\varepsilon^{5/2}),\\
\hat{\tau}^\text{SF}_{\cal ST} &= \frac{4\sqrt{2 \pi } G  \mathcal{H} k^{3/2} \left(c_2^* b_1-c_1^* b_2-b_2^* c_1+b_1^* c_2\right)}{a^2} \sqrt{\epsilon }\delta  +\cdots  
\quad\sim {\cal O} (\varepsilon^{5/2}),\\
\hat{\tau}^\text{CM}_{\cal ST} &=\frac{20\sqrt{2 \pi } G  \mathcal{H} k^{3/2} \left(c_2^* b_1-c_1^* b_2-b_2^* c_1+b_1^* c_2\right)}{a^2} \frac{1}{\sqrt{\epsilon}}  +\cdots  
\quad\sim {\cal O} (\varepsilon^{1/2}).
\end{align}
Here, ${\cal H}/k = -(k\eta)^{-1}$ $(=\sigma_S)$ in the expression is regarded small.
The contribution is a bit weaker than that of $\cal{T}$.
[For the Bunch-Davies initial conditions, the coefficients become
$c_1^* b_1+c_2^* b_2+b_1^* c_1+b_2^* c_2 = 2(c_2 b_2^* + c_2^* b_2)$ and
$c_2^* b_1-c_1^* b_2-b_2^* c_1+b_1^* c_2 = 2(c_2 b_2^* - c_2^* b_2)$.]

\section{Conclusions}
\label{Conc}

In this paper, we have systematically investigated the second-order effective energy-momentum tensor (2EMT) 
arising from the quadratic-order perturbations of scalar and tensor modes in the context of inflationary cosmology. 
By incorporating both scalar and tensor perturbations of the metric and their interaction with the scalar field (inflaton), 
we aimed to provide a comprehensive understanding of their collective back-reaction effects on the inflationary background
as a completion of the work in Ref.~\cite{Cho:2022maa}.

2EMT exhibits strong gauge dependence, primarily due to the scalar perturbation components.
Our analysis reveals several distinct behaviors for 2EMT across three primary gauge conditions: 
the longitudinal gauge, the spatially flat gauge, and the comoving gauge. 

The scalar-only contributions to 2EMT were found to dominate in the short-wavelength (high-momentum) limit ($k \gg {\cal H}$), 
where perturbations oscillate rapidly.
(The tensor-only contributions are equally significant in the $00$- and $33$-components.)
Conversely, in the long-wavelength (low-momentum) regime ($k \ll {\cal H}$), 
the tensor-only and coupled scalar-tensor components contribute more significantly to 2EMT. 
This result emphasizes that back-reaction effects from tensor modes persist even at large scales.
These findings underline the wavelength-dependent nature of 2EMT, 
and emphasize the need for careful gauge selection to accurately capture 
the effects of perturbative back-reaction in inflationary cosmology.

Findings in this work suggest that the scalar-tensor coupled terms, 
in particular in the long-wavelength limit,
could play an important role in shaping the effective stress-energy during inflation. 
The results imply that back-reaction effects from perturbations are not negligible,
and may impact the evolution of the inflationary background in subtle but significant ways. 
The off-diagonal energy-momentum tensor components resulting from these coupled terms
plays the role of shear viscosity,
which drives the spacetime to evolve {\it anisotropically} \cite{Cho:2022xku,Cho:2024jvn}.
Understanding these effects is essential for accurate predictions of inflationary outcomes, 
particularly with the improving precision of cosmological observations.

Future research could include extending this framework to exploring the scalar-tensor interaction in alternative gauges, 
and assessing observational signatures of 2EMT contributions,
such as their imprints on the cosmic microwave background and gravitational wave spectra.
At the current stage, however, it is not very plausible to discuss the observational implications 
because there still exist gauge issues; 2EMT depends on the gauge choices. 
Studying more gauge conditions might help in understanding the gauge issue in a relation to the observational implications.

It is also worthwhile to investigate the quantum effect in the future.
Involving with the tensor mode, there could be a quantum effect \cite{Polarski:1995jg} which was absent 
in the FRW universe with only the scalar-mode perturbation \cite{Cho:2020zbh}.

Our investigation can be extended to the post-inflationary (preheating) stage 
at which the scalar field exhibits the oscillatory behaviour.
A massive-scalar field ($V = m^2\phi^2/2$), for example, 
behaves as a pressureless dust in the oscillatory regime,
and the scale factor becomes $a(\eta) \propto \eta^2$.
The set-up will then be that of the perturbation of a scalar field in the Friedmann universe.
This will be a mixture of our previous~\cite{Cho:2020zbh,Cho:2022maa} and current work.
The perturbations produced at this stage have very short wavelengths,
and come back into the horizon at the early stage of the Friedmann universe.
The {\it curvature perturbation} does not grow much, 
and its effect is not significant to leave a signature in the {\it density perturbation}.

When the second scalar field is introduced, however, the situation may become interesting.
When another massive-scalar field is introduced, 
it will also play as a pressureless dust,
and the background universe will still settle down to a Friedmann universe.
When two fields are decoupled, they will not produce significant effects.
When two fields are coupled, however, the parametric resonance may induce a nontrivial effect 
such as the well-known ``nongaussianity" \cite{Enqvist:2004ey}. 
Regarding that the nongaussianity is a high-order effect, 
this set-up might induce an effect also in 2EMT that we consider.
It will be interesting to investigate 2EMT in this set-up in the future.

In principle, adding another scalar field to our set-up is straightforward.
However, the gauge invariant variables $\overline{\delta\phi_i}$ cannot be solved 
in a simple form by the Bardeen variable as was done for the single field in Eq.~\eqref{GIdeltaphi2}.
Therefore, it is not easy at once to expect what new effect will be brought to us by the second field,
in particular, when we consider a second-order effect such as 2EMT.
However, it will be a nice attempt to extended our work to the preheating stage with multiple-scalar fields.

\subsection*{Acknowledgements}

The author is grateful to Seung Hun Oh and Jinn-Ouk Gong for useful discussions,
and to Rajibul Shaikh for reading the draft.
This work was supported by the grants from the National Research Foundation
funded by the Korean government, 2020R1A2C1013266.

\end{document}